\documentclass{emulateapj}
\usepackage{amssymb}
\usepackage{amsmath}

\usepackage[]{natbib}
\usepackage[figuresright]{rotating}
\usepackage{color}

\graphicspath{{figures/}}

\newcommand{\fermi}{\emph{Fermi}}
\newcommand{\pass}[1]{{\texttt Pass~#1}}
\newcommand{\passsix}{\pass{6}}
\newcommand{\passseven}{\pass{7}}
\newcommand{\passeight}{\pass{8}}

\begin{document}

\title{New Fermi-LAT event reconstruction reveals more high-energy gamma rays from Gamma-ray bursts}

\author{
W.~B.~Atwood\altaffilmark{1}, 
L.~Baldini\altaffilmark{2}, 
J.~Bregeon\altaffilmark{3}, 
P.~Bruel\altaffilmark{4}, 
A.~Chekhtman\altaffilmark{5}, 
J.~Cohen-Tanugi\altaffilmark{6}, 
A.~Drlica-Wagner\altaffilmark{7}, 
J.~Granot\altaffilmark{8,9}, 
F.~Longo\altaffilmark{10,11}, 
N.~Omodei\altaffilmark{7,12}, 
M.~Pesce-Rollins\altaffilmark{3,13}, 
S.~Razzaque\altaffilmark{14}, 
L.~S.~Rochester\altaffilmark{7}, 
C.~Sgr\`o\altaffilmark{3}, 
M.~Tinivella\altaffilmark{3}, 
T.~L.~Usher\altaffilmark{7}, 
S.~Zimmer\altaffilmark{15,16}
}
\altaffiltext{1}{Santa Cruz Institute for Particle Physics, Department of Physics and Department of Astronomy and Astrophysics, University of California at Santa Cruz, Santa Cruz, CA 95064, USA}
\altaffiltext{2}{Universit\`a  di Pisa and Istituto Nazionale di Fisica Nucleare, Sezione di Pisa I-56127 Pisa, Italy}
\altaffiltext{3}{Istituto Nazionale di Fisica Nucleare, Sezione di Pisa, I-56127 Pisa, Italy}
\altaffiltext{4}{Laboratoire Leprince-Ringuet, \'Ecole polytechnique, CNRS/IN2P3, Palaiseau, France}
\altaffiltext{5}{Center for Earth Observing and Space Research, College of Science, George Mason University, Fairfax, VA 22030, resident at Naval Research Laboratory, Washington, DC 20375, USA}
\altaffiltext{6}{Laboratoire Univers et Particules de Montpellier, Universit\'e Montpellier 2, CNRS/IN2P3, Montpellier, France}
\altaffiltext{7}{W. W. Hansen Experimental Physics Laboratory, Kavli Institute for Particle Astrophysics and Cosmology, Department of Physics and SLAC National Accelerator Laboratory, Stanford University, Stanford, CA 94305, USA}
\altaffiltext{8}{Department of Natural Sciences, The Open University of Israel, 1 University Road, POB 808, Ra'anana 43537, Israel}
\altaffiltext{9}{email: j.granot@herts.ac.uk}
\altaffiltext{10}{Istituto Nazionale di Fisica Nucleare, Sezione di Trieste, I-34127 Trieste, Italy}
\altaffiltext{11}{Dipartimento di Fisica, Universit\`a di Trieste, I-34127 Trieste, Italy}
\altaffiltext{12}{email: nicola.omodei@stanford.edu}
\altaffiltext{13}{email: melissa.pesce.rollins@pi.infn.it}
\altaffiltext{14}{University of Johannesburg, Department of Physics, University of Johannesburg, Auckland Park 2006, South Africa, }
\altaffiltext{15}{Department of Physics, Stockholm University, AlbaNova, SE-106 91 Stockholm, Sweden}
\altaffiltext{16}{The Oskar Klein Centre for Cosmoparticle Physics, AlbaNova, SE-106 91 Stockholm, Sweden}

\begin{abstract}
Based on the experience gained during the four and a half years of the mission,
the \fermi-LAT collaboration has undertaken a comprehensive revision of the 
event-level analysis going under the name of \passeight. Although it is not yet
finalized, we can test the improvements in the new event reconstruction with 
the special case of the prompt phase of bright Gamma-Ray Bursts (GRBs), where 
the signal to noise ratio is large enough that loose selection cuts are sufficient to
identify gamma-rays associated with the source. Using the new event 
reconstruction, we have re-analyzed ten GRBs
previously detected by the LAT for which an x-ray/optical follow-up was
possible and found four new gamma rays with energies greater than 10 GeV in
addition to the seven previously known. Among these four is a 27.4 GeV 
gamma-ray from GRB~080916C, which has a redshift of 4.35, thus making it 
the gamma ray with 
the highest intrinsic energy ($\sim$147 GeV) detected from a GRB.
We present here the salient aspects of the new event reconstruction and
discuss the scientific implications of these new high-energy gamma rays, 
such as constraining extragalactic background light models, Lorentz invariance 
violation (LIV) tests, the prompt emission mechanism and the bulk Lorentz 
factor of the emitting region.

\end{abstract}

\keywords{Extragalactic Background Light, Fermi Large Area Telescope, Gamma-Ray Burst, Lorentz factor}

\section{Introduction}\label{sec:intro}

The Large Area Telescope (LAT) on-board the \fermi\ Gamma-Ray Space Telescope 
is a pair-conversion telescope designed to detect gamma rays above 
$\sim$20~MeV. The instrument is comprised of three subsystems acting in synergy
to identify and characterize gamma-ray interactions: a silicon 
tracker-converter
(TKR), a hodoscopic electromagnetic calorimeter (CAL) and a segmented 
anti-coincidence detector (ACD). By design, the TKR subsystem divides into 
two distinct sections: \emph{front} and
\emph{back}---the latter featuring six times thicker conversion foils.
Since these two sections are notably different from the standpoint of the
angular resolution and the contamination from misclassified cosmic rays, in 
the following we shall
analyze separately front- and back- converting candidate gamma rays. We refer 
the reader to~\citet{LATPaper} for further details on the LAT. 

Defining the event classes used for high-level scientific analysis is a 
complex process
involving many different steps: the event reconstruction, the adjudication of
the event energy and direction and the final event classification. In the
following we shall refer to this process as the \emph{event-level analysis}.
The pre-launch event-level analysis was solely based on Monte Carlo simulations
of the instrument performance and its particle environment---though it is worth
emphasizing that a significant effort was put into validating such
simulations \citep[see, e.g.:][]{baldini:190}. The event selection has been 
periodically updated to reflect the
constantly improving knowledge of the detector and the environment in which it
operates. \passseven\ \citep{P7Paper}, released in August 2011,
represents the latest major iteration of this incremental process. 

In parallel with the development of \passseven, the LAT collaboration has 
undertaken a coherent long-term effort to develop \passeight, aimed at an 
extensive revision of the entire event-level analysis~\citep{p8proceedings}. 
Recovering the effective area lost due to
residual signals from out-of-time cosmic-ray events (\emph{ghosts} hereafter),
was the original and main
motivation for this effort. As we shall see in the following, the scope of this
development has substantially expanded along the way.

The full event-level analysis for \passeight\ is currently under development. 
Therefore we cannot yet characterize its improvements in terms of instrument 
response functions. We can, however, test the improvements in the new 
event reconstruction by systematically searching for events not previously 
recognized as gamma rays during the prompt phases of bright Gamma-Ray Bursts
(GRBs) for which the signal to noise is large enough that loose selection cuts 
on quantities measured to classify events are sufficient to identify gamma rays
associated with the source.

In Section \ref{sec:recon} we briefly review some aspects of the LAT event
reconstruction, with emphasis on the modifications being introduced in \passeight.
In Sections~\ref{sec:selection} and \ref{sec:results} we discuss the analysis
underlying the search for new high-energy gamma rays and the results of this
search. Finally, in Section~\ref{sec:discussion} we discuss the implications of
these newly-found high-energy gamma rays.

\begin{deluxetable}{lcc}
  \tablewidth{\columnwidth}
  \tablecaption{GRBs used in this work}
  \tablehead{
    \colhead{GRB name} & \colhead{redshift} & \colhead{$\theta _{\rm LAT}$}} 
  \startdata
  GRB~080916C & 4.35 & $48{\fdg}8$\\ 
  GRB~090323  & 3.57 & $57{\fdg}2$\\
  GRB~090328  & 0.74 & $64{\fdg}6$\\
  GRB~090510  & 0.90 & $13{\fdg}6$\\
  GRB~090902B & 1.82 & $50{\fdg}8$\\
  GRB~090926  & 2.11 & $48{\fdg}1$\\
  GRB~091003  & 0.90 & $12{\fdg}3$\\
  GRB~091208B & 1.06 & $55{\fdg}6$\\
  GRB~100414A & 1.37 & $69{\fdg}0$\\
  GRB~110731A & 2.83 & $3{\fdg}4$\\[-6pt]
  \enddata
  \label{tab:grbs}
  \tablenotetext{}{The ten GRBs with measured redshift from the First Fermi GRB catalog~\citep{GRBCatalog} used in this analysis. The angle between the GRB and the LAT boresight is also listed in the last column.}
\end{deluxetable}

\section{Event reconstruction}\label{sec:recon}

A detailed description of the LAT event reconstruction is beyond the scope
of this paper. In the following we shall only give a brief description of
the development being implemented in the context of \passeight\ that is
relevant for the analysis presented here. We refer the reader
to~\citet{p8proceedings} for more details.

\subsection{Tracker reconstruction}\label{subsec:tkr_recon}

High-energy gamma-ray interactions in the CAL tend to generate \emph{backsplash}
in the lower portion of the TKR, i.e, randomly hit strips due to secondary
particles that have no relation to the trajectory of the original gamma ray.
For back-converting events, and especially at large incidence angle, it is
not uncommon for this backsplash to represent the vast majority of the TKR
hits.

The \passsix/\passseven\ TKR reconstruction is based on a track-by-track \emph{combinatoric}
pattern recognition---seeded by the CAL information when available.
As such, it is subject to confusion in backsplash-dominated events,
particularly if the initial position and direction estimates from the CAL are
not accurate. These features combine to produce two main effects: (i) the loss 
of events where the TKR reconstruction fails to find any tracks, and (ii) the 
migration of events from the core of the point-spread function (PSF) to the tails because of poorly 
reconstructed tracks.

In \passeight\ we introduced a \emph{global} approach, largely decoupled from the
CAL reconstruction, which looks at the gamma-ray conversion as a pre-shower
process and attempts to model this process by linking hits together into one or
more tree-like structures. The individual tracks are then extracted from these
structures and fitted. This new pattern recognition proved to be significantly
more efficient at finding tracks and more robust in terms of pointing accuracy.
Tests with Monte Carlo simulations and flight data show that the new TKR 
pattern recognition has the potential to provide a 15--20\% increase of the
high-energy acceptance, with even larger improvement in the off-axis effective
area, especially for photons converting in the lower part of the TKR.

\subsection{Calorimeter reconstruction}\label{subsec:cal_recon}

The \passsix/\passseven\ CAL reconstruction treats the energy deposit in the 
CAL as
a monolithic entity, grouping all the crystals with greater than 4 MeV energy 
deposited together. Residual ghost signals
in the CAL away from the gamma-ray shower can result in such a large lever arm
in the moments analysis used to derive the shower direction that they can
introduce substantial errors in the measurement of the centroid and direction
of the shower itself. Since the matching in event position and direction 
between the TKR and the CAL
constitutes an important input to the background rejection, this is actually
one of the main mechanisms for the ghost-induced loss of effective area at high
energy.

In \passeight\ we introduced a clustering stage, based on a Minimum Spanning Tree
algorithm, which proved to be effective in separating the genuine gamma-ray
signal from the ghost one. In addition, the 3D shower profile fit,
which is our primary energy reconstruction method at high energy, was
substantially improved. While the objective of this part of the work was to
extend the energy reach of the LAT above 1~TeV the new method
proved to provide an approximately 10\% improvement in the energy resolution 
over the entire energy and inclination angle range.

\subsection{ACD reconstruction}\label{subsec:acd_recon}

The basic purpose of the ACD reconstruction is to match tracks in the TKR and
hits in the ACD to find reasons to classify an event as a charged particle. 
The most significant improvement in
the ACD reconstruction we introduced in \passeight\ was to propagate the full
covariance matrices associated to the TKR tracks to the ACD surfaces---i.e.,
effectively we now measure the distances between tracks and ACD hits in terms
of measurement uncertainties rather than absolute lengths.

\begin{deluxetable*}{lccccccccc}
  \tablewidth{\textwidth}
  \tablecaption{Basic properties of the eleven gamma rays relevant for this work}
   \tablehead{ 
    \colhead{Run Id--Event Id} & \colhead{$t - T_0$ [s]} & 
    \colhead{Type} & \multicolumn{2}{c}{Energy [GeV]} &
    \multicolumn{2}{c}{Angle to source [$^\circ$]} &  \colhead{$p_{\rm all}$} & \colhead{photon class}\\
    \colhead{}&\colhead{} & \colhead{}& \colhead{\passsix} & \colhead{\passeight} & \colhead{\passsix} & \colhead{\passeight} & \colhead{} & \colhead{\passsix}}\\
\startdata
\multicolumn{10}{c}{GRB\,080916C ($\theta_{\rm LAT} = 48{\fdg}8$, $z = 4.3$)}\\
\hline
243215785--2033380 & 16.545 & back  & 13.2 & 12.4 & 0.09 &  0.11 & 1.000 & Diffuse\\
243215785--2075096$^\star$ & 40.509 & back  & - & 27.4 & - &  0.07 & 1.000 & - \\
\hline
\multicolumn{10}{c}{GRB\,090510 ($\theta_{\rm LAT} = 13{\fdg}6$, $z = 0.9$)}\\
\hline
263605997--3472705 & 0.828 & front  & 31.3 & 29.9 & 0.09 &  0.08 & 0.999 & Transient\\
\hline
\multicolumn{10}{c}{GRB\,090902B ($\theta_{\rm LAT} = 50{\fdg}8$, $z = 1.8$)}\\
\hline
273579835--4719473 & 11.671 & front  & 11.2 & 11.9 & 0.21 &  0.07 & 0.999 & Transient\\
273579835--4724519$^\star$ & 14.166 & back  & 14.2 & 14.2 & 2.61 &  0.11 & 0.980 & - \\
273579835--4748164$^\star$ & 26.168 & back  & - & 18.1 & - &  0.11 & 0.999 & - \\
273579835--4778868 & 42.374 & front  & 8.9 & 12.7 & 0.03 &  0.04 & 0.999 & Diffuse\\
273579835--4784978 & 45.608 & front  & 12.5 & 15.4 & 0.07 &  0.10 & 0.995 & Diffuse\\
273579835--4852498 & 81.746 & back  & 33.4 & 39.9 & 0.78 &  1.77 & 0.998 & Transient\\
\hline
\multicolumn{10}{c}{GRB\,090926 ($\theta_{\rm LAT} = 48{\fdg}1$, $z = 2.1$)}\\
\hline
275631595--173595 & 24.835 & front  & 19.6 & 19.5 & 0.05 &  0.09 & 0.999 & Diffuse\\
\hline
\multicolumn{10}{c}{GRB\,100414A ($\theta_{\rm LAT} = 69{\fdg}0$, $z = 1.4$)}\\
\hline
292903615--2268542$^\star$ & 33.365 & front  & 29.8 & 29.7 & 7.64 &  0.16 & 0.999 & - \\[-6pt]
    \enddata
    \label{tab:events_table}
  \tablenotetext{}{In the first column is run id and event id, following is 
    the arrival time, the conversion type, the
    reconstructed energy provided by \passsix\ and \passeight, the 
    reconstructed angle to the GRB provided by \passsix\ and \passeight, 
    the preliminary \passeight\ signal probability and the \passsix\ event class for the photons from the First LAT GRB catalog~\citep{GRBCatalog}. For each GRB, 
    $\theta_{\rm LAT}$ indicates the angle between the GRB and the LAT 
    boresight and $z$ is the redshift from ~\citep{GRBCatalog}.
    The four gamma-ray candidates recovered owing to the reconstruction 
    improvements provided by \passeight\ are indicated by the $^\star$ symbol 
    in the first column. The \passeight\ angle to source for event 4852498 of GRB\,090902B falls marginally outside our ROI. This is addressed in section \ref{sec:selection}.}
\end{deluxetable*}

\section{Data selection}\label{sec:selection}

Among the bursts in the First LAT GRB catalog~\citep{GRBCatalog}, we concentrate on the ten GRBs for which an x-ray/optical follow-up (and therefore a measurement of the redshift) was obtained (see Table \ref{tab:grbs} for a listing of the GRBs used). The typical localization error for these bursts is negligible compared with the event-by-event direction accuracy of the LAT and, for all practical purposes, we can consider the localizations measured in the optical or X-ray afterglows as the true source positions when defining the region of interest (ROI). We further refine our sample by considering only energies greater than 10~GeV, where the width of the core of the LAT PSF is close to its asymptotic high-energy limit. It is important to stress here that analysis described in~\citet{GRBCatalog} was done using the \passsix\ version of the gamma-ray data (and the associated \texttt{P6\_V3\_TRANSIENT} IRFs), rather than the reprocessed \passseven\ data made available in August 2011. However, the event reconstruction between \passsix\ and \passseven\ remained essentially unchanged and therefore this is largely irrelevant for the purpose of this paper.

For each GRB we reprocessed all the available data
within 90~s from the trigger time using the \passeight\ event reconstruction
available at the time of writing. We select gamma-ray candidates by 
requiring that the reconstruction find at least one track and that this track 
extrapolates to more than 4 radiation
lengths of active material in the CAL (this helps avoiding poorly
reconstructed events). In addition, we use the ACD to remove
\emph{likely} charged-particle events by requiring that the track-tile
association most likely to veto the event is incompatible with being generated
by a minimum ionizing particle. Note that the event selection used here does 
not include any requirement on the quality of the direction/energy 
reconstruction.

The choice of the ROI is dictated by the need to minimize the amount of solid 
angle over which we integrate the background of residual (mis-classified) 
cosmic rays while at the same time retaining a reasonable efficiency
for well-reconstructed gamma rays. For each GRB we used a circular ROI around
the nominal source position with a radius of $0\fdg6$ for front-converting
and $1\fdg2$ for back-converting events. It is worth noting that, while
these are comparable with the PSF 95\% contaiment radii for the cleanest
\passsix/\passseven\ event classes, based on Monte Carlo simulations we 
estimate that the actual containment level for the
back-converting events passing our loose selection cuts is only about 80\%, so
that the ROI cut has a significant impact on the event topology of our sample
(i.e., it plays a role in selecting well-reconstructed events).

The expected rate of background events passing these basic selection criteria
can be estimated from the flight data by sideband subtraction\citep{P7Paper} 
using an annulus
around the source position and rescaling the number of counts to the solid
angle subtended by the original ROI. As the level of charged-particle
background varies across the \fermi\ orbit, the results are slightly different
for each individual GRB, but on average we expect $\sim 0.1$ cosmic-ray events
passing our basic selection cuts within each of the 90~s time windows.

\section{Results}\label{sec:results}

The First LAT GRB catalog includes seven candidate gamma rays with energies 
greater than 10~GeV
associated with the ten GRBs considered here; in the reprocessed version
of the data we find four additional (previously misreconstructed) events
passing our selection criteria. In Table~\ref{tab:events_table} we summarize 
the basic properties of the four new gamma rays together with the seven 
previously known ones.
All of the seven aforementioned gamma-ray candidates pass our loose selection
cuts (and their topologies are highly gamma-ray-like).
However one of them (a $\sim33$~GeV event from GRB~090902B) is reconstructed
as being marginally outside our ROI. While this is not entirely surprising
(the quality of the direction reconstruction for this particular event is
fairly poor both in the \passsix\ and in the \passeight\ versions of the
event-level analysis), assessing the actual probability for this event to be
associated with the GRB in the context of any of the actual \passeight\ event
classes will only be possible once the definitions of the classes are frozen 
and the corresponding response functions defined.

It is interesting to note how the mechanisms through which these events are
recovered tie to the problematic aspects of the LAT event reconstruction
outlined in Section~\ref{sec:recon}: two of them had no tracks in \passsix,
one was significantly mis-tracked (to more than~$7^\circ$ off the source) and
the last one was compromised by a ghost cluster in the CAL. We would like to stress here that the \passeight\ event reconstruction is in its final phase; therefore we are confident that the basic topological properties of these four new photons will not vary significantly and that no additional photons will be found in the same 90 s interval explored in this work.

Table~\ref{tab:events_table} shows how the preliminary \passeight\ event
selection described in~\citet{p8proceedings} assigns a fairly high value for 
the measure of gamma-ray probability $p_{\rm all}$ to all the eleven gamma-ray 
candidates considered here. While this is an uncalibrated quantity with no 
\emph{direct} physical meaning
(and it might very well change in future iterations of the \passeight\
event-level analysis) 
we find that the fraction of background events with such a high value of
$p_{\rm all}$ is of the order of $10^{-3}$. This is effectively a multiplicative
factor for the $\sim 0.1$~Hz rate of background events quoted in
Section~\ref{sec:selection}. 

Finally, we note that most of the events in the table have angular distances
to the nominal source position much smaller than the radius
of our ROI. Under the reasonable assumption that the
background is approximately isotropic in our $\sim 1^{\circ}$ circle, we would
expect background events to be preferentially near the edge of the ROI
(just because it subtends a larger solid angle).

We estimated the increase in effective area over the \passsix\ \texttt{TRANSIENT} class expected for our selection%
\footnote{We included the effect of the ROI in our estimate by requiring, in
  addition to the preliminary Pass 8 event selection that we applied, that the 
  angle between the true and reconstructed gamma-ray direction is smaller than 
  the radius of the ROI itself.}
by means of a gamma-ray Monte Carlo simulation similar to those that
we routinely use for generating effective area tables. Between 10 and 30~GeV
this improvement is of the order of $\sim 10\%$ at $50^\circ$ off-axis angle
and reaches $\sim 50\%$ at $70^\circ$. Note that the choice of off-axis angles 
corresponds to those of the GRBs from which we recovered the four new gamma 
rays. Though the small statistics in our GRB event sample does not allow a 
validation of the increase
in effective area, these factors are consistent with our findings. 
We also stress that the factors refer to the particular selection used 
in this analysis and do not represent the actual performance of any of the 
still-forthcoming \passeight\ event classes. 

\subsection{Spectral analysis}

\begin{deluxetable*}{lcccc}
\tablewidth{\textwidth}
\tablecaption{Spectral analysis results}
\tablehead{
\colhead{} & \colhead{GRB\,080916C} & \multicolumn{2}{c}{GRB\,090902B}& GRB\,100414A
}
\startdata
N$_{obs}$: Number of \passsix\ events & & & \\ 
with energy $>$ new \passeight\ event & 0 & \multicolumn{2}{c}{1} & 0 \\
\hline
Best fit value of the gamma ray index & 2.1 $\pm$ 0.09 &   \multicolumn{2}{c}{1.96 $\pm$ 0.07}& 2.1 $\pm$ 0.4 \\
Expected number of events (N$_{exp}$) &  0.5 &  3.3 & 2.5 &  0.1 \\
Probability of observing N$_{obs}$ events   &  0.60  & 0.13 & 0.21 & 0.88 \\
Probability of observing $>$N$_{obs}$ events  &  0.40 & 0.63$^{\ast}$ & 0.71 & 0.12 \\[-6pt]
\enddata
\label{tab:events}
\tablenotetext{}{Output from the spectral analysis for the three GRBs from which we 
  recover the four new gamma-rays. The best fit value of the gamma ray index are taken from \citet{GRBCatalog}. The value with the $^{\ast}$ is the 
  probability of observing two additional events and refers to event 4724519 
  from GRB090902B.}
\end{deluxetable*}

As stated in Section \ref{sec:intro}, a final \passeight\ event-level analysis is not yet available and therefore we cannot perform a spectral analysis. We can however estimate the probability to detect these high-energy gamma rays, given the spectral properties inferred from \passsix\ analysis. Using \texttt{gtobssim}, we simulated 90~s observations of a very bright source located at the position of each of the three GRBs with new candidate gamma rays, using the best fitted values (from \citet{GRBCatalog}) for the index of the power-law spectrum. We normalized the output of the simulation to the observed number of \passsix\ \texttt{TRANSIENT} counts above 100~MeV in a ROI of 5$^{\circ}$ in order to estimate the expected number of events, $N_{exp}$, above the energy of the new \passeight\ gamma ray. Finally, we use the Poisson distribution to compute the probability of observing exactly $N_{obs}$ events when the number of expected events is $N_{exp}$.
In addition, we calculate the probabilities of observing at least one additional gamma ray -- or two additional gamma rays for GRB\,090902B -- with an energy equal to or greater than those recovered with \passeight. Results are reported in Table \ref{tab:events}.

We have studied the potential impacts that a spectral evolution during the considered time interval (90 s) may have on the resulting probabilities, P(N$_{exp}$,N$_{obs}$) by repeating the Monte Carlo simulation with a varying index for the spectral distribution of gamma rays. We find that the associated variation in the probability in the worst case\footnote{GRB\,080916C~\citep{GRB080916C} is the GRB in our sample that shows the largest spectral variation, with a gamma ray index varying from 2.3 to 2.1 with an error of 0.09.} is of the order of 10\%--15\%. 
The calculated probabilities suggest that the additional gamma rays are statistically consistent with the shape and intensity of the spectra derived using \passsix\ data.

\section{Discussion}\label{sec:discussion}
Our most interesting finding is the $27.4\;$GeV gamma ray from GRB\,080916C
that was detected $40.5\;$s after the burst onset. At a redshift of $z
\approx 4.35$ the measured energy corresponds to an energy of $\approx
147\;$GeV in the GRB cosmological rest frame. This is the
highest intrinsic energy measured so far for a gamma ray from a GRB. 

\begin{figure*}[!htb*]
  \includegraphics[width=0.5\linewidth]{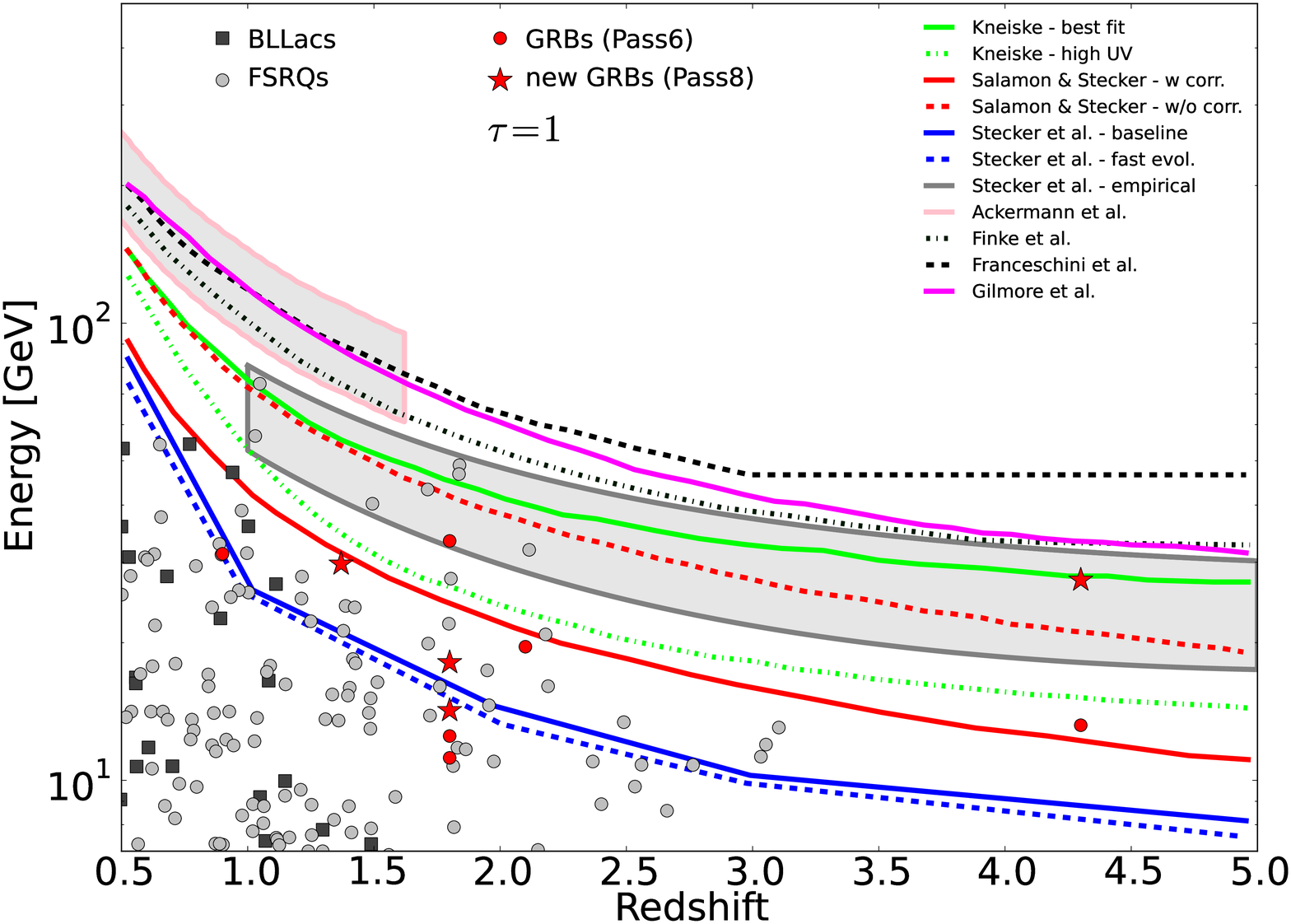}
  \includegraphics[width=0.5\linewidth]{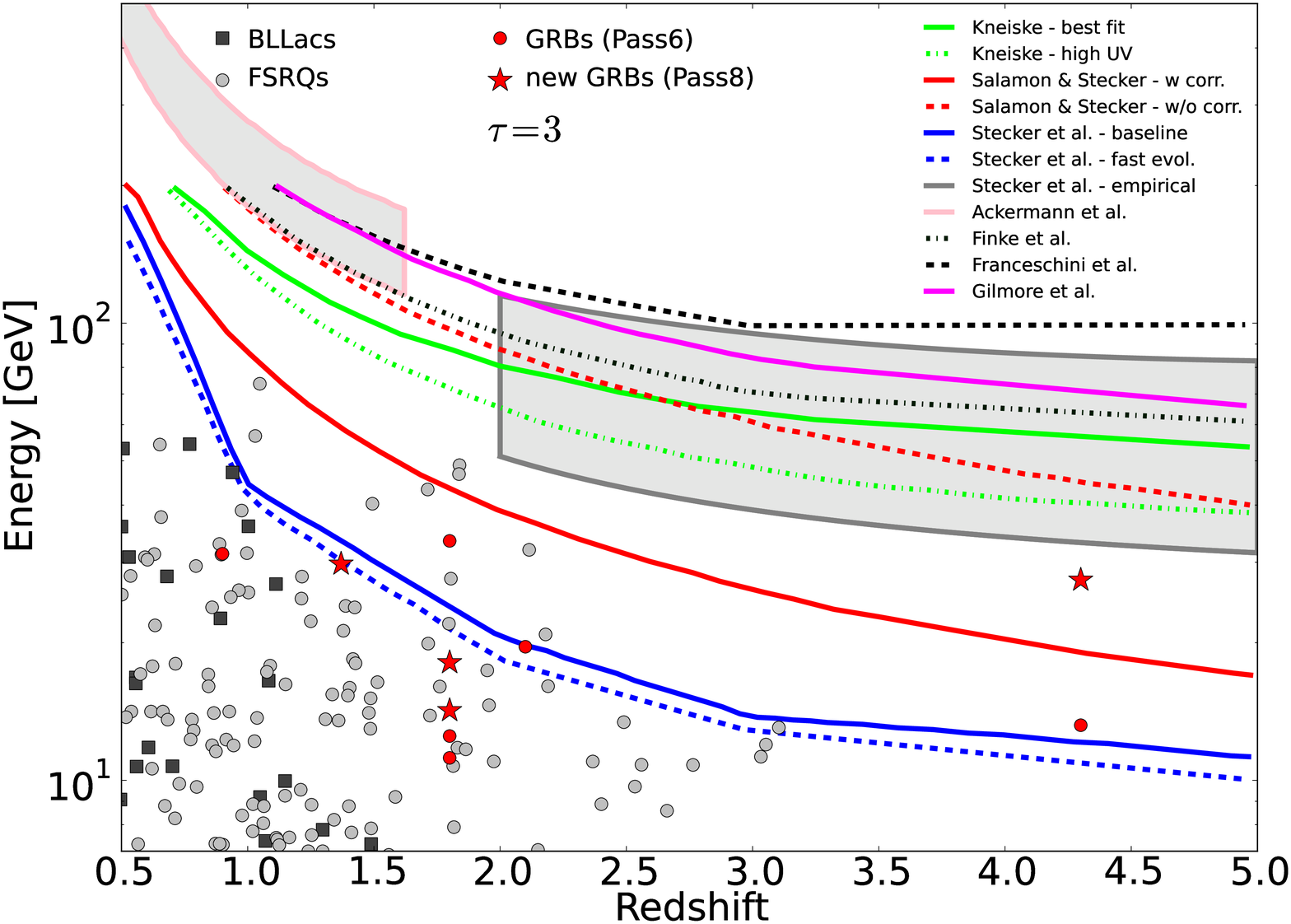}
  \caption{Highest-energy gamma rays from blazars~\citep{2010ApJ...723.1082A},
    and GRBs~\citep{GRBCatalog} seen by the LAT. 
    Predictions of optical depth due to pair production, $\tau_{\gamma\gamma}$=1 (left panel) and
    $\tau_{\gamma\gamma}$=3 (right panel) from various EBL models are indicated
    by lines. The shaded area outlined in grey is the prediction from 
    \citet{0004-637X-761-2-128} and the shaded area outlined in pink is 
    the best fit value (1$\sigma$) measured by 
    \citet{Ackermann:2012sza}. Gamma rays above model predictions in this 
    figure traverse an EBL medium with
    a high gamma-ray opacity. The four new gamma rays presented in this work
    are represented by the red stars.}
  \label{fig:ebl}
\end{figure*}

The high energy of the new gamma ray from GRB~080916C is very constraining for 
a possible origin from synchrotron radiation. 
A reasonable assumption for the acceleration time of the radiating electron, 
that it is at least the time
it takes to complete one Larmor gyration\footnote{This assumption implies a maximum electron Lorentz
factor of $\gamma_{\rm max} =\sqrt{3e/\sigma_T B'}$ where $B'$ is the
comoving (measured in the rest frame of the emitting plasma) magnetic
field. The corresponding comoving typical synchrotron gamma ray energy
averaged over an isotropic pitch-angle distribution is $E'_{\rm
syn,max} = 3heB'\gamma_{\rm max}^2/(16 m_ec) = (27/64)m_ec^2/\alpha$,
where $\alpha = e^2/\hbar c\approx 1/137$ is the fine structure
constant. The corresponding observed energy is $E_{\rm syn,max} =
E'_{\rm syn,max}\Gamma/(1+z)\approx 29.5(1+z)^{-1}(\Gamma/1000)\;$GeV.
Therefore, a synchrotron origin for this gamma ray would imply $E_{\rm
obs} \lesssim E_{\rm syn,max}$ or $\Gamma \gtrsim 2030 [(1+z)/3] (E_{\rm
obs}/20\,{\rm GeV})$.  The peak of the electron synchrotron spectral
emissivity is $0.29$ times the value for $E_{\rm syn,max}$ used above, and 
using it would
increase the limit correspondingly (by a factor of 1/0.29) to $\Gamma
\gtrsim 7000[(1+z)/3](E_{\rm obs}/20\,{\rm GeV})$. Allowing the
acceleration time to be as small as the time it takes to deflect the
electron by one radian (which is quite extreme) lowers the limit by a
factor of $2\pi$, to $\Gamma \gtrsim 323[(1+z)/3](E_{\rm obs}/20\,{\rm
GeV})$.  Combining such a short acceleration time with the factor of
$0.29$ mentioned above leads to $\Gamma \gtrsim 1110[(1+z)/3](E_{\rm
obs}/20\,{\rm GeV})$.
Recently, \citet{2012MNRAS.tmpL.529K} have proposed a way to lower this limit by assuming two zones with significantly 
different magnetic field strength, where in the lower-field region electrons can be accelerated to high Lorentz factors, 
and then radiate energetic synchrotron gamma rays after crossing to the high-field region. This could in principle 
accommodate the production of $\gtrsim 100\;$GeV gamma rays with significantly lower bulk Lorentz factors.},
would imply a minimum bulk Lorentz factor of the emitting region larger than
$5000$. 
Similarly, the
$29.7\;$GeV gamma ray from GRB~100414A at $z = 1.37$ would require
$\Gamma\gtrsim 2300$ for a synchrotron origin. For 
GRB~090902B~\citep{GRB090902B} at $z =1.822$ the two new gamma rays are less constraining 
than the $33.4\;$GeV
gamma ray detected $82\;$s after the burst onset with \passsix, after the end of the
prompt emission, which implies $\Gamma\gtrsim 3200$ (making a
synchrotron origin unlikely for the $33.4\;$GeV gamma ray due to its later arrival time).

Given the arrival time of $40.5\;$s after the burst onset of the $27.4\;$GeV 
gamma ray from GRB\,080916C, during interval d defined in ~\citep{GRB080916C}, 
the lower limit on $\Gamma$ due to 
intrinsic opacity to pair production is increased by only $15\%$
compared to the limit from the $13\;$GeV gamma ray observed in the same
time interval, of $\Gamma_{\rm min}\approx 600$ for a 
simple one-zone model, or $\sim 3$ times lower than this for a more realistic
self-consistent time-dependent model
\citep{2008ApJ...677...92G, Hascoet:2011gp}.

Due to its later arrival time, the constraints that the new gamma ray from 
GRB\,080916C
provides on linear ($n=1$) Lorentz invariance violation (LIV) are
slightly weaker (by 15\%) than the previously highest
energy gamma ray from the same GRB (of energy $\approx 13.2\;$GeV
detected at $t = 16.5\;$s after the GRB trigger time, which
implied\footnote{The limits we quote conservatively use the lowest
values within the 1-$\sigma$ confidence intervals for the gamma ray
energy (and the GRB redshift when relevant).} $\xi_1 = M_{\rm
QG,1}/M_{\rm Planck} > 0.11$). For a quadratic leading LIV term ($n=2$)
it does slightly better with $M_{\rm QG,2} > 1.13\times 10^{10}\;{\rm
GeV}/c^2$, which is only $\approx 2.6$ times below the best limit from
GRB~090510~\citep{2009Natur.462..331A}. The limits from the
other new gamma rays are not as constraining.

A very interesting implication arises for the extragalactic background
light (EBL), from the fact that a 27.4~GeV gamma ray has reached
us from a fairly high redshift of $z \approx 4.35$, and was not
attenuated (through pair production, $\gamma\gamma \to e^+e^-$) by the
EBL. In particular, it is useful to
compare the constraints that it provides to those from previously-detected 
gamma rays from GRBs~\citep{GRBCatalog} and AGN~\citep{2010ApJ...723.1082A}, as
illustrated in Figure~\ref{fig:ebl}. It is the most constraining
gamma ray so far from a GRB (see Figs. 3 and 5 in \citet{2010ApJ...723.1082A};
notice in particular that Fig. 5 also applies to the
newly-found $27.4\;$GeV from GRB\,080916C). Moreover, it is in fact comparable 
to or even slightly more constraining than the \fermi-LAT gamma rays from AGN 
(for most EBL models\footnote{
A description of the different models is beyond the scope of this work; we refer the reader to the original works on the various EBL models
\citep[e.g.,][]{1998ApJ...493..547S,2006ApJ...648..774S,0004-637X-761-2-128,2002A&A...386....1K,2004A&A...413..807K,2005AIPC..745...23P,
2009MNRAS.399.1694G,2008A&A...487..837F,2009ApJ...697..483R,2010ApJ...712..238F}} especially for $\tau = 3$ as shown in Figure~\ref{fig:ebl}).

In conclusion, the improvements in event reconstruction implemented 
in \passeight\ promise to yield scientific gains, as illustrated in this work.  

\acknowledgments
The \textit{Fermi} LAT Collaboration acknowledges generous ongoing support
from a number of agencies and institutes that have supported both the
development and the operation of the LAT as well as scientific data analysis.
These include the National Aeronautics and Space Administration and the
Department of Energy in the United States, the Commissariat \`a l'Energie Atomique
and the Centre National de la Recherche Scientifique / Institut National de Physique
Nucl\'eaire et de Physique des Particules in France, the Agenzia Spaziale Italiana
and the Istituto Nazionale di Fisica Nucleare in Italy, the Ministry of Education,
Culture, Sports, Science and Technology (MEXT), High Energy Accelerator Research
Organization (KEK) and Japan Aerospace Exploration Agency (JAXA) in Japan, and
the K.~A.~Wallenberg Foundation, the Swedish Research Council and the
Swedish National Space Board in Sweden.

Additional support for science analysis during the operations phase is gratefully
acknowledged from the Istituto Nazionale di Astrofisica in Italy and the Centre National d'\'Etudes Spatiales in France.

\bibliographystyle{aa}

\bibliography{pesc0504}

\end{document}